\documentclass[submission,copyright,creativecommons]{eptcs}
\providecommand{\event}{EXPRESS 2011}

\usepackage{stmaryrd}
\usepackage{amsmath}
\usepackage{amssymb}
\usepackage{mathrsfs}
\usepackage{amsthm}
\usepackage{extarrows}
\usepackage{xspace}
\usepackage{tikz}
	\usetikzlibrary{arrows,shapes,automata,backgrounds,petri}

\usepackage{SynchronyCausalityPi}

\title{Synchrony vs Causality\\ in the Asynchronous Pi-Calculus%
\thanks{This work was supported by the DFG (German Research Foundation), grants {NE-1505/2-1} and {GO-671/6-1}.}}
\author{Kirstin Peters
\institute{School of EECS, TU Berlin, Germany}
\email{kirstin.peters@tu-berlin.de}
\and
Jens-Wolfhard Schicke
\institute{Institute for Programming and Reactive Systems, TU Braunschweig, Germany}
\email{drahflow@gmx.de}
\and
Uwe Nestmann
\institute{School of EECS, TU Berlin, Germany}
\email{uwe.nestmann@tu-berlin.de}
}
\def\titlerunning{Synchrony vs Causality in the Asynchronous Pi-Calculus}
\def\authorrunning{K. Peters, J.-W. Schicke \& U. Nestmann}

\begin{document}

\maketitle

\begin{abstract}
We study the relation between process calculi that differ in their either synchronous or asynchronous interaction mechanism.  Concretely, we are interested in the conditions under which synchronous interaction can be implemented using just asynchronous interactions in the $\pi$-calculus.  We assume a number of minimal conditions referring to the work of Gorla: a ``good'' encoding must be compositional and preserve and reflect computations, deadlocks, divergence, and success.  Under these conditions, we show that it is \emph{not} possible to encode synchronous interactions without introducing additional causal dependencies in the translation.

\end{abstract}

\section{Introduction}

We study the relation between process calculi that differ in their either synchronous or asynchronous interaction mechanism. Synchronous and a\-syn\-chro\-nous interactions are the two basic paradigms of interactions in distributed systems. While synchronous interactions are widely used in specification languages, asynchronous interactions are often better suited to implement real systems. We are interested in the conditions under which synchronous interactions can be implemented using just asynchronous interactions, i.e., in the conditions under which it is possible to encode the synchronous $ \pi $-calculus into its asynchronous variant. To partially answer this question, we examine the role of causality for encoding synchrony.

Of course, we are not interested in trivial or meaningless encodings. Instead we consider only those encodings that ensure that the original term and its encoding show to some extent the same abstract behaviour. Unfortunately, there is no consensus about what properties make an encoding ``good'' (compare e.g.~\cite{parrow08}). Instead, we find separation results as well as encodability results with respect to very different conditions, which naturally leads to incomparable results. Among these conditions, a widely used criterion is \emph{full abstraction}, i.e. the preservation and reflection of equivalences associated to the two compared languages. There are lots of different equivalences in the range of $ \pi $-calculus variants. Since full abstraction depends, by definition, strongly on the chosen equivalences, a variation in the respective choice may change an encodability result into a separation result, or vice versa. Unfortunately, there is neither a common agreement about what kinds of equivalence are well suited for language comparison|again, the results are often incomparable. To overcome these problems, and to form a more robust and uniform approach for language comparison, Gorla \cite{gorla08,gorla10} identifies five criteria as being well suited for separation as well as encodability results. In this paper, we rely on these five criteria. Compositionality and name invariance stipulate structural conditions on a good encoding. Operational correspondence requires that a good encoding preserves and reflects the computations of a source term. Divergence reflection states that a good encoding shall not exhibit divergent behaviour, unless it was already present in the source term. Finally, success sensitiveness requires that a source term and its encoding have exactly the same potential to reach successful state.

A discussion on synchrony versus asynchrony cannot be separated from a discussion of choice.  When processes communicate via message-passing along channels, they do not only listen to one channel at a time|they concurrently listen to a whole selection of channels.  Choice operators just make this natural intuition explicit; moreover, their mutual exclusion property allows us to concisely describe the particular effect of message-passing actions on the process's local state.  Asynchronous send actions make no sense as part of a mutually exclusive selection, as they cannot be prevented from happening.  Consequently, the asynchronous calculus only offers input-guarded choice.  In contrast, synchronous send actions also allow for the definition of mixed choice: selections of both input and output actions.

It is well known that there is a good encoding from the choice-free synchronous \piCal-calculus into its asynchronous variant \cite{boudol92,hondaTokoro91,honda:soundness}. It is also well-known \cite{palamidessi03,gorla10,petersNestmann10} that there is no good encoding from the full $ \pi $-calculus|the synchronous \piCal-calculus including mixed choice|into its asynchronous variant if the encoding translates the parallel operator homomorphically. Palamidessi was the first to point out that mixed choice strictly raises the absolute expressive power of the synchronous \piCal-calculus compared to its asynchronous variant. Analysing this result \cite{petersNestmann10}, we observe that it boils down to the fact that the full \piCal-calculus can break merely syntactic symmetries, where its asynchronous variant can not; there is no need to refer to a semantic problem like the existence of solutions to leader election.  Moreover, as already Gorla \cite{gorla10} states, the  condition of homomorphic translation of the parallel operator is rather strict. Therefore, Gorla proposes the weaker criterion of compositional translation of the source language operators (see Definition \ref{def:compositionality} at page \pageref{def:compositionality}). As claimed in \cite{petersNestmann11}, this weakening of the structural condition on the encoding of the parallel operator turns the separation result into an encodability result, i.e., there is an encoding from the synchronous $ \pi $-calculus (including mixed choice) into its asynchronous variant with respect to the criteria of Gorla\footnote{Note that this encoding is neither prompt nor is the assumed equivalence $ \asymp $ strict, so the separation results of \cite{gorla08} and \cite{gorla10} do not apply here.}. Analysing the encoding attempt given in \cite{petersNestmann11}, we observe that it introduces additional causal dependencies, i.e., causal dependencies that were not present in the source term and thus introduced by the encoding function. Note, that a step $ B $ is considered causally dependent on a previous step $ A $, if $ B $ depends on the availability of data produced by $ A $. In this paper, we show that this is a general phenomenon of encoding synchrony.

Thus, as the main contribution of this paper, we show that|in the asynchronous $\pi$-calculus|there is a strong connection between synchronous interactions and causal dependencies. More precisely, we show that it is \emph{not} possible to encode synchronous interactions within a completely asynchronous framework without introducing additional causal dependencies in the translation. Moreover, we discuss the role of mixed choice to derive this result for the $ \pi $-calculus. 
The companion paper \cite{schickePetersGoltz11} presents a similar result in the context of Petri nets. Hence, this connection between synchronous interactions and causal dependencies is presumably no effect of the representation of concurrent systems in either the $ \pi $-calculus or Petri nets, but rather a phenomenon of synchronous and asynchronous interactions in general. 

\paragraph{Overview of the Paper.} 

In \S\ref{sec:technicalPreliminaries}, we introduce the synchronous $ \pi $-calculus and its asynchronous variant. We revisit some notions and results of \cite{petersNestmann10}, recall the five criteria of \cite{gorla10} to measure the quality of an encoding, and talk about a definition of causality. In \S\ref{sec:synchronyCausality}, we present our separation result and discuss some observations on its proof. We conclude in \S\ref{sec:conclusion} with a comparison to a similar result in \cite{schickePetersGoltz11}.

\section{Technical Preliminaries} \label{sec:technicalPreliminaries}

\subsection{The \piCal-calculus} \label{sec:piCalculus}

Our source language is the monadic $ \pi $-calculus as described for instance in \cite{sangiorgiWalker01}. Since the main reason for the absolute difference in the expressiveness of the full $ \pi $-calculus compared to the asynchronous $ \pi $-calculus is the power of mixed choice we denote the full $ \pi $-calculus also by \piMix.

Let $ \names $ denote a countably infinite set of names and $ \coNames $ the set of co-names, i.e., $ \coNames = \Set{ \Out{n} \mid n \in \names} $. We use lower case letters $ a, a', a_1, \ldots, x, y, \ldots $ to range over names.

\begin{definition}[\piMix] \label{def:piMix}
  The set of process terms of the \emph{$ \pi $-calculus (with mixed choice)}, denoted by $ \piMixProc $, is given by
	\begin{align*}
		P & \;\mathop{::=}\;
                \RestrictedTerm{n}{P}
                \sep P_1 \mid P_2
                \sep !P
                \sep \sum_{i \in \indexSet} \guard_i.P_i 
	\end{align*}
	where $ \guard \;\mathop{::=}\; \Input{y}{x} \; \mid \; \Output{y}{z} $ for some names $ n, x, y, z \in \names $ and a finite index set $ \indexSet $.
\end{definition}
\noindent
The interpretation of the defined process terms is as usual. Since all examples and counterexamples within this paper are CCS-like we omit the objects of actions. Moreover we denote the empty sum with $ \nullTerm $ and omit it in continuations. As usual we often notate a sum $ \sum_{i \in \Set{ i_1, \ldots, i_n }} \guard_i.P_i $ by $ \guard_{i_1}.P_{i_1} + \ldots + \guard_{i_n}.P_{i_n} $.

As target language, we use \piAsyn, the asynchronous \piCal-calculus (see \cite{hondaTokoro91} or \cite{boudol92}).

\begin{definition}[\piAsyn] \label{def:piAsyn}
  The set of process terms of the \emph{asynchronous $ \pi $-calculus}, denoted by $ \piAsynProc $, is given by
  \begin{align*}
    P \;\mathop{::=}\;
    & \RestrictedTerm{n}{P} \sep P_1 \mid P_2 
    \sep !P
    \sep \nullTerm
    \sep \Output{y}{z} 
    \sep \Input{y}{x}.P
    \sep \Match{a}{b}P
 \end{align*}
  for some names $ n, a, b, x, y, z \in \names $.
\end{definition}
\noindent

Here, we equip the target language with the match operator, because it is used in \cite{petersNestmann11}, which introduces the only good encoding that we are aware of between the synchronous $ \pi $-calculus (with mixed choice) and its asynchronous variant. With \cite{petersNestmann11}, this encoding depends on the availability of the match operator in the target language; we do not yet know whether there is such an encoding without match. Since matching do increase the expressive power of the asynchronous \piCal-calculus (see \cite{carboneMaffeis03}), answering this question is an important task for future work. However, note that the proof of our main result in Theorem \ref{thm:encodingSynchronyIntroduceCausality} does not depend on this decision. 

As shown by the encoding in \cite{nestmann00} one could also use separate choice within an asynchronous variant of the calculus without a significant effect on its expressive power. We claim, that our main result does not depend on the decision whether to allow separate choice or not. In Section \ref{sec:theorem} we give some hints on how to change the proof to capture separate choice in the target language.

We use capital letters $ P, P', P_1, \ldots, Q, R, \ldots $ to range over processes. If we refer to processes without further requirements, we denote elements of $ \piMixProc $; we sometimes use just $ \proc $ when the discussion applies to both calculi. Let $ \FreeNames{P} $ denote the set of \emph{free names} in $ P $. Let $ \BoundNames{P} $ denote the set of \emph{bound names} in $ P $. Likewise, $ \Names{P} $ denotes the set of all \emph{names} occurring in $ P $. Their definitions are completely standard.

\begin{figure}[ht]
	\begin{align*}
		\begin{array}{|c|}
			\hline
			\\
			P \equiv Q \begin{aligned}[t]
					& \text{ if } Q \text{ can be obtained from } P \text{ by renaming one or more of the bound names in P,}\\
					& \text{ silently avoiding name clashes }
				\end{aligned}\\
			\\
			\quad P \mid \nullTerm \equiv P \hspace*{2em} P \mid Q \equiv Q \mid P \hspace*{2em} P \mid \left( Q \mid R \right) \equiv \left( P \mid Q \right) \mid R \hspace*{2em} \Match{a}{a}P \equiv P \hspace*{2em} !P \equiv P \mid !P \quad \\
			\\
			\RestrictedTerm{n}{\nullTerm} \equiv \nullTerm \hspace*{2em} \RestrictedTerm{n}{\RestrictedTerm{m}{P}} \equiv \RestrictedTerm{m}{\RestrictedTerm{n}{P}} \hspace*{2em} P \mid \RestrictedTerm{n}{Q} \equiv \RestrictedTerm{n}{\left( P \mid Q \right)} \text{ if } n \notin \FreeNames{P}\\
			\\
			\hline
		\end{array}
	\end{align*}
	\caption{Structural Congruence.} \label{fig:SC}
\end{figure}

The \emph{reduction semantics} of \piMix and \piAsyn are jointly given by the transition rules in Figure \ref{fig:RSPiMix}, where \emph{structural congruence}, denoted by $ \equiv $, is given by the rules in Figure \ref{fig:SC}. Note that the rule $ \textsc{Com}_{\operatorname{a}} $ for communication in \piAsyn is a simplified version of the rule $ \textsc{Com}_{\operatorname{mix}} $ for communication in \piMix. The differences between these two rules result from the differences in the syntax, i.e. the lack of choice and the fact that only input can be used as guard in \piAsyn. As usual, we use $ \equivAlpha $ if we refer to alpha-conversion (the first rule of Figure \ref{fig:SC}) only.

\begin{figure}[ht]
	\begin{align*}
		\begin{array}{|c|}
			\hline
			\\
			\quad \textsc{Com}_{\operatorname{mix}} \quad \left( \ldots + \Input{y}{x}.P + \ldots \right) \mid \left( \ldots + \Output{y}{z}.Q + \ldots \right) \longmapsto \Set{ \Subst{z}{x} }P \mid Q \quad\\
			\\
			\textsc{Com}_{\operatorname{a}} \quad \Input{y}{x}.P \mid \Output{y}{z} \longmapsto \Set{ \Subst{z}{x} }P\\
			\\
			\textsc{Par} \quad \dfrac{P \longmapsto P'}{P \mid Q \longmapsto P' \mid Q} \hspace*{3em} \textsc{Res} \quad \dfrac{P \longmapsto P'}{\RestrictedTerm{n}{P} \longmapsto \RestrictedTerm{n}{P'}}\\
			\\
			\textsc{Cong} \quad \dfrac{P \equiv P' \quad P' \longmapsto Q' \quad Q' \equiv Q}{P \longmapsto Q}\\
			\\
			\hline
		\end{array}
	\end{align*}
	\caption{Reduction Semantics of \piMix and \piAsyn.} \label{fig:RSPiMix}
\end{figure}

We use $ \sigma $, $ \sigma' $, $ \sigma_1 $, \ldots to range over substitutions. A substitution is a mapping $ \Set{ \Subst{x_1}{y_1}, \ldots, \Subst{x_n}{y_n} } $ from names to names. The application of a substitution on a term $ \Set{ \Subst{x_1}{y_1}, \ldots, \Subst{x_n}{y_n} }\left( P \right) $ is defined as the result of simultaneously replacing all free occurrences of $ y_i $ by $ x_i $ for $ i \in \Set{ 1, \ldots, n } $, possibly applying alpha-conversion to avoid capture or name clashes. For all names $ \names \setminus \Set{ y_1, \ldots, y_n } $ the substitution behaves as the identity mapping. Let $ \id $ denote identity, i.e. $ \id $ is the empty substitution.

Let $ P \longmapsto $ ($ P \not\longmapsto $) denote existence (non-existence) of a step from $ P $, i.e. there is (no) $ P' \in \proc $ such that $ P \longmapsto P' $. Moreover, let $ \Longmapsto $ be the reflexive and transitive closure of $ \longmapsto $ and let $ \longmapsto^{\omega} $ define an infinite sequence of reduction steps.

The first quality criteria presented in Section \ref{sec:qualityEncoding} is compositionality. It induces the definition of a context parametrised on a set of names for each operator of \piMix. A context $ \mathcal{C}\left( \hole_1, \ldots, \hole_n \right) $ is simply a $ \pi $-term, i.e. a \piAsyn-term in case of Definition \ref{def:compositionality}, with $ n $ holes. Putting some \piAsyn-terms $ P_1, \ldots, P_n $ in this order into the holes $ \hole_1, \ldots, \hole_n $ of the context, respectively, gives a term denoted $ \mathcal{C}\left( P_1, \ldots, P_n \right) $. Note that a context may bind some free names of $ P_1, \ldots, P_n $. The arity of a context is the number of its holes.

\subsection{Symmetry in the \piCal-calculus}

A \emph{network} of degree $ n $ is a process $ \RestrictedTerm{\tilde{x}}{\left( P_1 \mid \ldots \mid P_n \right)} $ for some $ n \in \nat $, some $ P_1, \ldots, P_n \in \proc $ and a sequence of names $ \tilde{x} $. We refer to $ P_1, \ldots, P_n $ as the processes of the network. Note that the processes of a network can be networks itself. A \emph{symmetric network} of degree $ n $ is a network of degree $ n $ such that $ P_i = \sigma^{i - 1}\left( P_1 \right) $ for all $ i \in \Set{ 2, \ldots, n } $ and some substitution $ \sigma $, called \emph{symmetry relation}, such that $ \sigma^n = \id $. The minimal degree of a symmetry relation $ \sigma $ is the smallest $ n > 0 $ such that $ \sigma^n = \id $. A \emph{symmetric execution} is an execution starting at a symmetric network of degree $ n $, returning to a symmetric network of the same degree after any $ n $'th step, and which is either infinite or terminates in a symmetric network of degree $ n $. Let \piSep be the subcalculus of \piMix with separate but no mixed choice, i.e. there is no sum with both input- and output-guarded summands, and let $ \piSepProc $ be its set of processes. The first and the last author \cite{petersNestmann10} prove that it is not possible in \piSep to break the symmetry of a symmetric network. More precisely, in Theorem 4.4 in \cite{petersNestmann10}, it is shown that any symmetric network in $ \piSepProc $ has at least one symmetric execution. Since $ \piAsynProc $ is a subset of $ \piSepProc $, we obtain the following lemma.

\begin{lemma} \label{lem:symmetricExecution}
	Every symmetric network in $ \piAsynProc $ has at least one symmetric execution.
\end{lemma}

Moreover, by Lemma 5.4 in \cite{petersNestmann10}, we know that if the minimal degree of the symmetry relation is smaller than the degree of the symmetric network, then not only the network can be subdivided into a network of symmetric networks but also its symmetric execution. As already done in \cite{petersNestmann10}, we use this Lemma in the context of a symmetric network $ P \mid P $ for some arbitrary $ P \in \piAsynProc $. $ P \mid P $ is a symmetric network of degree $ 2 $ with the symmetry relation $ \id $. The minimal degree of $ \id $ is $ 1 $. Let $ P \mid P \Longmapsto \RestrictedTerm{\tilde{x}}{\left( P' \mid \sigma\left( P' \right) \right)} $ be a symmetric execution of $ P \mid P $ for some $ P' \in \piAsynProc $, a sequence of names $ \tilde{x} $, and some symmetry relation $ \sigma $ with $ \sigma^2 = \id $. By Lemma 5.4 in \cite{petersNestmann10}, this symmetric execution can be subdivided such that there is an execution $ P \Longmapsto \RestrictedTerm{\tilde{x}'}{P'} $ for some sequence of names $ \tilde{x}' $.

\begin{lemma} \label{lem:subdividingSymmetricExecution}
	Let $ P, P' \in \piAsynProc $, $ \tilde{x} $ be a sequence of names, and $ \sigma $ be a symmetry relation with $ \sigma^2 = \id $. Every symmetric execution $ P \mid P \Longmapsto \RestrictedTerm{\tilde{x}}{\left( P' \mid \sigma\left( P' \right) \right)} $ can be subdivided such that $ P \Longmapsto \RestrictedTerm{\tilde{x}'}{P'} $ for some sequence of names $ \tilde{x}' $.
\end{lemma}

\subsection{Quality Criteria for Encodings} \label{sec:qualityEncoding}

Gorla presented in \cite{gorla10} a small framework of five criteria well suited for language comparison. We use this five criteria to measure the quality of an encoding $ \encoding $ from \piMix into \piAsyn, i.e. an encoding $ \encoding $ is ``good'' if it fulfils the five criteria proposed by Gorla. Note that for the definition of these criteria a behavioural equivalence $ \asymp $ on the target language is assumed. Its purpose is to describe the abstract behaviour of a target process, where abstract basically means with respect to the behaviour of the source term.

The five conditions are divided into two structural and three semantic criteria. The structural criteria include (1) \emph{compositionality} and (2) \emph{name invariance}. The semantic criteria include (3) \emph{operational correspondence}, (4) \emph{divergence reflection} and (5) \emph{success sensitiveness}. In the following we use $ S, S', S_1, \ldots $ to range over terms of the source language and $ T, T', T_1, \ldots $ to range over terms of the target language.

Intuitively, an encoding is compositional if the translation of an operator depends only on the translation of its parameters.  To mediate between the translations of the parameters the encoding defines a unique context for each operator, whose arity is the arity of the operator. Moreover, the context can be parametrised on the free names of the corresponding source term. Note that our result is independent of this parametrisation.

\begin{definition}[Criterion 1: Compositionality] \label{def:compositionality}
	The encoding $ \encoding $ is \emph{compositional} if, for every k-ary operator $ \mathbf{op} $ of $ \piMix $ and for every subset of names $ N $, there exists a k-ary context $ \Context{N}{\mathbf{op}}{\hole_1, \ldots , \hole_k} $ such that, for all $ S_1, \ldots, S_k $ with $ \FreeNames{S_1} \cup \ldots \cup \FreeNames{S_k} = N $, it holds that
	\begin{align*}
		\Encoding{\mathbf{op}\left( S_1, \ldots, S_k \right)} = \Context{N}{\mathbf{op}}{\Encoding{S_1}, \ldots , \Encoding{S_k}}.
	\end{align*}
\end{definition}

The second structural criterion states that the encoding should not depend on specific names used in the source term. Of course, an encoding that translates each name to itself simply preserves this condition. However, it is sometimes necessary and meaningful to translate a name into a sequence of names or to reserve a couple of names for the encoding, i.e. to give them a special function within the encoding. To ensure that there are no conflicts between the names used by the encoding function for special purposes and the source term names, the encoding is enriched with a renaming policy $ \renamingPolicy $, i.e., a substitution from names into sequences of names\footnote{To keep distinct names distinct Gorla assumes that $ \forall n, m \in \names \logdot n \neq m \text{ implies } \RenamingPolicy{n} \cap \RenamingPolicy{m} = \emptyset $, where $ \RenamingPolicy{x} $ is simply considered as set here.}. Based on such a renaming policy an encoding is independent of specific names if it preserves all substitutions $ \sigma $ on source terms by a substitution $ \sigma' $ on target terms such that $ \sigma' $ respects the changes made by the renaming policy.

\begin{definition}[Criterion 2: Name Invariance] \label{def:nameInvariance}
	The encoding $ \encoding $ is \emph{name invariant} if, for every $ S $ and $ \sigma $, it holds that
	\begin{align*}
		\Encoding{\sigma\left( S \right)} \begin{cases} \equivAlpha \sigma'\left( \Encoding{S} \right) & \quad \text{if } \sigma \text{ is injective}\\ \asymp \sigma'\left( \Encoding{S} \right) & \quad \text{otherwise} \end{cases}
	\end{align*}
	where $ \sigma' $ is such that $ \RenamingPolicy{\sigma\left( a \right)} = \sigma'\left( \RenamingPolicy{a} \right) $ for every $ a \in \names $.
\end{definition}

The first semantic criterion is operational correspondence, which consists of a soundness and a completeness condition. \emph{Completeness} requires that every computation of a source term can be simulated by its translation, i.e., the translation does not reduce the computations of the source term. \emph{Soundness} requires that every computation of a target term corresponds to some computation of the corresponding source term, i.e., the translation does not introduce new computations.

\begin{definition}[Criterion 3: Operational Correspondence] \label{def:operationalCorrespondence}
	The encoding $ \encoding $ is \emph{operationally corresponding} if it is
	
	\begin{tabular}{ll}
		\emph{Complete}: & for all $ S \Longmapsto S' $, it holds that $ \Encoding{S} \Longmapsto \asymp \Encoding{S'} $;\\
		\emph{Sound}: & for all $ \Encoding{S} \Longmapsto T $, there exists an $ S' $ such that $ S \Longmapsto S' $ and $ T \Longmapsto \asymp \Encoding{S'} $.
	\end{tabular}
\end{definition}
\noindent
Note that the Definition of operational correspondence relies on the equivalence $ \asymp $ to get rid of junk possibly left over within computations of target terms. Sometimes, we refer to the completeness criterion of operational correspondence as operational completeness and, accordingly, for the soundness criterion as operational soundness.

The next criterion concerns the role of infinite computations in encodings.

\begin{definition}[Criterion 4: Divergence Reflection] \label{def:divergenceReflection}
  The encoding $ \encoding $ reflects divergence if, for every $ S $, $ \Encoding{S} \longmapsto^{\omega} $ implies $ S \longmapsto^{\omega} $.
\end{definition}

The last criterion links the behaviour of source terms to the behaviour of target terms.
With Gorla \cite{gorla10}, we assume a \emph{success} operator $ \success $ to be part of the syntax of both the source and the target language.  Likewise, we add $ \success $ to the syntax of \piMix in Definition \ref{def:piMix} and of \piAsyn in Definition \ref{def:piAsyn}. Since $ \success $ can not be further reduced, the operational semantics is left unchanged in both cases. Moreover, note that $ \Names{\success} = \FreeNames{\success} = \BoundNames{\success} = \emptyset $, so also interplay of  $\success$ with the $\equiv$-rules is smooth and does not require explicit treatment. The test for reachability of success is standard.

\begin{definition}[Success] \label{def:success}
	A process $ P \in \proc $ \emph{may lead to success}, denoted as $ P \Downarrow $, if (and only if) it is reducible to a process containing a top-level unguarded occurrence of $ \success $, i.e. $ \exists P', P'' \in \proc \logdot P \Longmapsto P' \wedge P' \equiv P'' \mid \success $.
\end{definition}
\noindent
Note that we choose may-testing here. However, as we claim, our main result in Theorem \ref{thm:encodingSynchronyIntroduceCausality} holds for must-testing, as well.

Finally, an encoding preserves the behaviour of the source term if it and its corresponding target term answer the tests for success in exactly the same way.

\begin{definition}[Criterion 5: Success Sensitiveness] \label{def:succesSensitiveness}
  The encoding $ \encoding $ is \emph{success sensitive} if, for every $ S $, $ S \Downarrow $ if and only if $ \Encoding{S} \Downarrow $.
\end{definition}
\noindent
Note that this criterion only links the behaviours of source terms and their literal translations but not of their continuations. To do so, Gorla relates success sensitiveness and operational correspondence by requiring that the equivalence on the target language never relates two processes $ P $ and $ Q $ such that $ P \Downarrow $ and $ Q \not\Downarrow $.

\begin{definition}[Success Respecting] \label{def:successRespecting}
	$ \asymp \; \subseteq \piAsynProc \times \piAsynProc $ is \emph{success respecting} if, for every $ P $ and $ Q $ with $ P \Downarrow $ and $ Q \not\Downarrow $, it holds that $ P \not\asymp Q $.
\end{definition}

\subsection{Causality}

Analysing the five criteria of the last section, we observe that there are two structural criteria to ensure that a good encoding is implementable, i.e., is of practical interest, and there are three criteria to ensure that the encoding preserves and reflects the main behaviour of source terms. However, there is no criterion requiring the preservation or reflection of causal dependencies.

For the \piCal-calculus usually two kinds of causal dependencies are distinguished (see \cite{priami96, borealeSangiorgi98}). The first one, called structural or subject dependencies, originates from the nesting of prefixes, i.e., from the structure of processes. A typical example of such a dependency is given by $ \RestrictedTerm{b}{\left( \Out{a}.\Out{b} \mid \In{b}.\Out{c} \right)} \mid \In{a} \mid \In{c} \longmapsto \RestrictedTerm{b}{\left( \Out{b} \mid \In{b}.\Out{c} \right)} \mid \In{c} \longmapsto \Out{c} \mid \In{c} \longmapsto \nullTerm $. The second step on channel $ b $ is causally dependent on the first step, because it unguards $ \Out{b} $. So $ b $ is causally dependent on $ a $. Similarly, $ c $ is causally dependent on $ b $, and by transitivity $ c $ is causally dependent on $ a $. The other kind of dependencies are called link or object dependencies and originate from the binding mechanisms on names. Here a typical example is $ \RestrictedTerm{x}{\left( \Output{y}{x} \mid \Out{x} \right)} $. In a labelled semantics the output on $ x $ is causally dependent on the extrusion of $ x $ by an output on $ y $, i.e. $ x $ is causally dependent on $ y $.

We observe that causal dependencies are defined as a condition between actions or names of actions. In the context of encodings this view is problematic, because steps are often translated into sequences of steps and  names may be translated into sequences of names. Moreover a sequence of steps simulating a single source term step may be interleaved with another such sequence or some target term steps used to prepare the simulation of another source term step, whose simulation may never be completed. So, what precisely does it mean for an encoding to preserve or respect causal dependencies? If source term names are translated into sequences of names should one consider the causal dependencies between all such translated names or only between some of them? Moreover how should an encoding handle names reserved for some special purposes of the encoding function, i.e., target term names that do not result from the translation of a source term name?

We have no final answer to these questions yet. However, in the next section we prove a separation result, which does not require a thorough answer to the questions above. Instead, we use a definition of causal dependencies that is based only on direct subject dependencies. So within this paper, a step $ B $ is considered causally dependent on a previous step $ A $, if $ B $ depends on the availability of a capability produced by $ A $. More precisely, step $ B $ is causally dependent on step $ A $, if $ A $ unguards some capability, i.e., some input or output prefix, which is consumed by step $ B $. An encoding preserves causal dependencies, if for any causal dependency between two steps of the source term there is a causal dependency between some steps of their simulations, and an encoding reflects causal dependencies, if for any causal dependency between two steps of different, completed simulations there is a causal dependency of the corresponding source term steps.

\section{Synchrony vs Causality} \label{sec:synchronyCausality}

In this section, we show that any good encoding from \piMix into \piAsyn introduces causal dependencies, i.e., is not causality respecting.

\subsection{Encoding Synchrony introduces Causal Dependencies} \label{sec:theorem}

To prove our main result we analyse the context introduced to encode the parallel operator and examine how this context has to interact with the encodings of its parameters to allow for a simulation of a source term step. We start with some observations concerning the three process terms
\begin{align*}
	P \deff \overline{a} + a \mid \overline{b} + b.\success, \quad Q \deff P \mid P, \quad \text{and} \quad R \deff \overline{a} + b + b.\success \mid \overline{b} + a + a.\success,
\end{align*}
which are used in the following lemmata as counterexamples. Note that we choose $ Q $, and $ R $ such that each of them is a symmetric network of degree $ 2 $ with either $ \sigma = \Set{ a/b, b/a } $ or $ \id $ as symmetry relation. Moreover to fix the context used to encode the parallel operator we choose $ P $, $ Q $, and $ R $ such that $ \FreeNames{P} = \FreeNames{Q} = \FreeNames{R} = \Set{ a, b } $. Hence by compositionality for each of these three terms the outermost parallel operator is translated by exactly the same context $ \ContextABPar{\hole_1, \hole_2} $.
\begin{obs}
	There exists a context $ \ContextABPar{\hole_1, \hole_2} $ such that $ \Encoding{P} = \ContextABPar{\Encoding{\overline{a} + a}, \Encoding{\overline{b} + b.\success}} $, $ \Encoding{Q} = \ContextABPar{\Encoding{P}, \Encoding{P}} $, and $ \Encoding{R} = \ContextABPar{\Encoding{\overline{a} + b + b.\success}, \Encoding{\overline{b} + a + a.\success}} $. \label{obs:PQR}
\end{obs}
We choose $ P $ such that none of its executions lead to success, i.e., $ P \not\longmapsto $ and $ P \not\Downarrow $. $ P \not\longmapsto $ implies, by operational soundness, that $ \Encoding{P} $ can not perform a step that changes its state modulo $ \asymp $, i.e. $ \Encoding{P} \Longmapsto T_P $ implies $ T_P \Longmapsto \asymp \Encoding{P} $ for all $ T_P \in \piAsynProc $. By success sensitiveness, $ P \not\Downarrow $ implies $ \Encoding{P} \not\Downarrow $. Because of that and since $ \asymp $ is success respecting we have $ T_P \not\Downarrow $ for all $ T_P \in \piAsynProc $ such that $ \Encoding{P} \Longmapsto T_P $.
\begin{obs}
	$ \forall T_P \in \piAsynProc \logdot \Encoding{P} \Longmapsto T_P \text{ implies } T_P \not\Downarrow $ \label{obs:P}
\end{obs}
Hence, any occurrence of $ \success$|if there is any|in the context $ \ContextABPar{\hole_1, \hole_2} $ is input guarded (since \piAsyn forbids output guards) and the context can not remove such a guard on its own.
In opposite to $ P $ we choose $ Q $ such that $ Q $ reaches an unguarded occurrence of success in any of its executions, i.e. $ Q \Longmapsto Q' $ implies $ Q' \Downarrow $ for all $ Q' \in \piMixProc $. By operational completeness, any execution $ Q \longmapsto Q_1 \longmapsto Q_2 \not\longmapsto $ of $ Q $ can be simulated by its encoding, i.e. $ \Encoding{Q} \Longmapsto Q_1' $, $ \Encoding{Q} \Longmapsto Q_2' $, and $ \Encoding{Q_1} \Longmapsto Q_2'' $, where $ Q_i' \asymp \Encoding{Q_i} $ for $ i \in \Set{ 1, 2 } $ and $ Q_2'' \asymp \Encoding{Q_2} $. Note that any (maximal) execution of $ Q $ is such that $ Q \longmapsto Q_1 \longmapsto Q_2 \not\longmapsto $ for some $ Q_1, Q_2 \in \piMixProc $. By operational soundness for each $ T_Q \in \piAsynProc $ such that $ \Encoding{Q} \Longmapsto T_Q $, there is some $ Q' \in \piMixProc $ such that $ Q \Longmapsto Q' $ and $ T_Q \Longmapsto \asymp \Encoding{Q'} $, i.e. there is some $ T_Q' \in \piAsynProc $ such that $ T_Q \Longmapsto T_Q' $ and $ T_Q' \asymp \Encoding{Q'} $. By success sensitiveness, $ \Encoding{Q'} \Downarrow $ and since $ \asymp $ is success respecting, we have $ T_Q' \Downarrow $. Thus, by Definition \ref{def:success}, we have $ T_Q \Downarrow $ for all $ T_Q \in \piAsynProc $ with $ \Encoding{Q} \Longmapsto T_Q $.
\begin{obs}
	$ \forall T_Q \in \piAsynProc \logdot \Encoding{Q} \Longmapsto T_Q \text{ implies } T_Q \Downarrow $ \label{obs:Q}
\end{obs}
At last we choose $ R $ such that some of its executions lead to success while some do not. $ R $ can reduce either to $ \success $ or to $ \mathbf{0} $. By operational completeness, $ \Encoding{R} $ can simulate both steps, i.e. $ \Encoding{R} \Longmapsto \asymp \Encoding{\success} $ and $ \Encoding{R} \Longmapsto \asymp \Encoding{\mathbf{0}} $. Since $ \Encoding{\success} \Downarrow $ and $ \Encoding{\mathbf{0}} \not\Downarrow $, and since $ \asymp $ is success respecting, we have $ \Encoding{\success} \not\asymp \Encoding{\mathbf{0}} $. By operational soundness, for all $ T_R \in \piAsynProc $ such that $ \Encoding{R} \Longmapsto T_R $ there is some $ R' \in \piMixProc $ such that $ R \Longmapsto R' $ and $ T_R \Longmapsto \asymp \Encoding{R'} $.
\begin{obs}
	$ \exists T_{R, 1}, T_{R, 2} \in \piAsynProc \logdot \Encoding{R} \Longmapsto T_{R, 1} \wedge \Encoding{R} \Longmapsto T_{R, 2} \wedge T_{R, 1} \Downarrow \wedge \, T_{R, 2} \not\Downarrow $ \label{obs:R}
\end{obs}
Our last observation concerns the structure of the context $ \ContextABPar{\hole_1, \hole_2} $. Because there is no choice operator in $ \piAsynProc $, the context $ \ContextABPar{\hole_1, \hole_2} $ has to place its parameters in parallel, as this is the only binary operator for processes. However, even if we allow separate choice in the target language the encodings of the parameters have to be placed in parallel, because placing them within a choice would not allow to use the encodings of both parameters to simulate target term steps. Consequently, there must be some $ \piAsynProc $-contexts $ \ContextA{\hole} $, $ \ContextB{\hole} $, $ \ContextC{\hole} $ with $ \Encoding{S_1 \mid S_2} \equiv \ContextABPar{\Encoding{S_1}, \Encoding{S_2}} \equiv \ContextA{\ContextB{\Encoding{S_1}} \mid \ContextC{\Encoding{S_2}}} $, for all source terms $ S_1, S_2 \in \piMixProc $ with $ \FreeNames{S_1 \mid S_2} = \Set{ a, b } $.
\begin{obs}
	$ \exists \ContextA{\hole}, \ContextB{\hole}, \ContextC{\hole} \logdot \ContextABPar{\hole_1, \hole_2} \equiv \ContextA{\ContextB{\hole_1} \mid \ContextC{\hole_2}} $ \label{obs:context}
\end{obs}

Learning from the separation result in \cite{petersNestmann10}, we know that any good encoding from \piMix into \piAsyn must break source term symmetries. To do so, we show that the context introduced by the encoding of the parallel operator (which is allowed in weakly compositional as opposed to homomorphic translations) must interact with the encodings of its parameters.
\begin{lemma}
	To simulate a source term step, $ \ContextABPar{\hole_1, \hole_2} $ and the encodings of its parameters have to interact. \label{lem:contextInteract}
\end{lemma}
Intuitively we show, that if there is no such interaction, then since $ Q $ is a symmetric network its encoding behaves as a symmetric network again. Since any execution of $ \Encoding{Q} $ leads to an unguarded occurrence of success, by symmetry and by Lemma \ref{lem:subdividingSymmetricExecution} there is an execution of $ \Encoding{P} $ leading to an unguarded occurrence of success, which contradicts Observation \ref{obs:P}.
\begin{proof}
	Assume the opposite, i.e. assume the context $ \ContextABPar{\hole_1, \hole_2} $ is such that possibly after some preprocessing steps of the context on its own, e.g. to unguard the parameters, the source term steps can be simulated without any interaction with the context. In this case, we have
	\begin{align*}
		\Encoding{Q} \stackrel{\ref{obs:PQR}}{=} \ContextABPar{\Encoding{P} \mid \Encoding{P}} \stackrel{\ref{obs:context}}{\equiv} \ContextA{\ContextB{\Encoding{P}} \mid \ContextC{\Encoding{P}}} \Longmapsto \RestrictedTerm{\tilde{y}}{\left( \sigma_1\left( \Encoding{P} \right) \mid \sigma_2\left( \Encoding{P} \right) \mid T_C \right)}
	\end{align*}
	for some constant term $ T_C $, a sequence of names $ \tilde{y} $, and two substitutions $ \sigma_1 $ and $ \sigma_2 $. Note that $ \sigma_1 $ and $ \sigma_2 $ capture renaming done by alpha conversion possibly necessary to pull restriction outwards. Since there is no need for an interaction, i.e. for a communication, with $ T_C $ to simulate source term steps, we can ignore it.
	
	If $ \sigma_1 = \sigma_2 $, then since these substitutions result from alpha conversion $ \sigma_1 = \sigma_2 = \id $. Then $ \Encoding{P} \mid \Encoding{P} $ is a symmetric network of degree $ 2 $ with $ \id $ as symmetry relation. By Lemma \ref{lem:symmetricExecution}, $ \Encoding{P} \mid \Encoding{P} $ has a symmetric execution. By Observation \ref{obs:Q}, $ \Encoding{Q} $ reaches an unguarded occurrence of success in any of its executions. Since the context and with it $ T_C $ can not reach success on its own and there is no interaction, $ \Encoding{P} \mid \Encoding{P} $ reaches success in its symmetric execution. Then there is some $ T_Q'' \in \piAsynProc $ such that $ \Encoding{P} \mid \Encoding{P} \Longmapsto \left( \nu \tilde{x} \right) \left( T_Q'' \mid \sigma_3\left( T_Q'' \right) \right) $ is a symmetric execution for some sequence of names $ \tilde{x} $ and some symmetry relation $ \sigma_3 $ of degree $ 2 $ and $ \left( \nu \tilde{x} \right) \left( T_Q'' \mid \sigma_3\left( T_Q'' \right) \right) $ has an unguarded occurrence of success. By symmetry and since $ \mathsf{n}\left( \success \right) = \emptyset $, this implies that $ T_Q'' $ as well as $ \sigma\left( T_Q'' \right) $ has an unguarded occurrence of success. Since $ 2 $ is not the minimal degree of identity, by Lemma \ref{lem:subdividingSymmetricExecution}, this symmetric execution can be subdivided such that $ \Encoding{P} \Longmapsto \left( \nu \tilde{x}' \right) T_Q'' $ for some sequence of names $ \tilde{x}' $. Then $ \Encoding{P} \Downarrow $, because of the unguarded occurrence of $ \success $ in $ T_Q'' $. That contradicts Observation \ref{obs:P}.
	
	The argumentation for $ \sigma_1 \neq \sigma_2 $ is similar but more difficult. In this case $ \sigma_1\left( \Encoding{P} \right) \mid \sigma_2\left( \Encoding{P} \right) $ is still a symmetric network whose symmetric execution leads to an unguarded occurrence of success. But since its symmetry relation is not $ \id $ we can not apply Lemma \ref{lem:subdividingSymmetricExecution}. However, because $ \sigma_1 $ and $ \sigma_2 $ result from alpha conversion, they rename free names of $ \Encoding{P} $ to fresh names. If $ \sigma_1\left( \Encoding{P} \right) $ and $ \sigma_2\left( \Encoding{P} \right) $ want to interact on such a fresh name, then they have first to exchange this fresh name over a channel known to both. Let us denote this channel by $ z $. So either $ \sigma_1\left( \Encoding{P} \right) $ receives a fresh name from $ \sigma_2\left( \Encoding{P} \right) $ over $ z $ or vice versa. By symmetry both terms have an unguarded input as well as an unguarded output on $ z $, so|instead of a communication between these two processes|$ \sigma_1\left( \Encoding{P} \right) $ can as well reduce on its own. Adding this argumentation to the argumentation in the proof of Lemma \ref{lem:subdividingSymmetricExecution} in \cite{petersNestmann10} we can prove again that the symmetric execution of $ \sigma_1\left( \Encoding{P} \right) \mid \sigma_2\left( \Encoding{P} \right) $ can be subdivided such that $ \sigma_1\left( \Encoding{P} \right) \Downarrow $. Because $ \Names{\success} = \emptyset $, this implies $ \Encoding{P} \Downarrow $.
	Hence, to simulate a source term step, the context necessarily has to interact with its parameters.
\end{proof}

Note that the only possibility for the context to interact with its parameters is by communication. So the context contains at least one capability, i.e., input or output prefix, that needs to be consumed to simulate a source term step. Without loss of generality let us assume that indeed only a single capability needs to be consumed to simulate a step, i.e. a single communication step of the context with (one of) its parameters suffices to enable the simulation of a source term step. The argumentation for a couple of necessary communication steps is similar. Let us denote this capability by $ \mu $\footnote{In case of a sequence of necessary steps, choose $ \mu $ such that it denotes the capability consumed at last in this sequence. In case there are different ways to enable the simulation of a step, consider a set of those capabilities with one $ \mu_i $ for each such way.}.

Next we show, that it is not possible to simulate two different source term steps between the parameters of $ \ContextABPar{\hole_1, \hole_2} $ at the same time.
\begin{lemma}
	At most one simulation of a source term step can be enabled concurrently by the context $ \ContextABPar{\hole_1, \hole_2} $. \label{lem:singleSimulation}
\end{lemma}
Here we use $ R $ as a counterexample. $ R $ can reduce either to $ \nullTerm $ or $ \success $, so the choice operator introduces mutual exclusion. Without choice mutual exclusion is not that easy to implement, because of its ability to immediately block an alternative reduction. We show that the simulation of these blocking introduces either deadlock or divergence.
\begin{proof}
	Assume the opposite, i.e. assume that the context $ \ContextABPar{\hole_1, \hole_2} $ provides several instances of $ \mu $, e.g. by replication, and with them it enables the simulation of different alternative source term steps concurrently. Consider the source term $ R $. Since $ \left\{ a/b, b/a \right\}\left( \overline{a} + b + b.\success \right) = \overline{b} + a + a.\success $, by name invariance, there is some substitution $ \sigma' $ such that $ \sigma'\left( \Encoding{\overline{a} + b + b.\success} \right) \equivAlpha \Encoding{\overline{b} + a + a.\success} $, i.e. these two terms are equal except to some renamings of free names.
	
	Note that $ R $ can perform either a step on channel $ a $ or $ b $. Since there is no choice operator in \piAsyn, the encodings of the capabilities of the sums $ \overline{a} + b + b.\success $ and $ \overline{b} + a + a.\success $ have to be placed somehow in parallel such that the simulation of the source term step on $ a $ does not immediately withdraw the encodings of the capabilities on $ b $ and vice versa. Thus, since the simulation of both steps of $ R $ are enabled concurrently, there is some point in the simulation of one source term step that disables the completion of the simulation of the respective other source term step. Therefore, one simulation has to consume some capability that is necessary to complete the other simulation. Remember, that we assume that the only capability of the context $ \ContextABPar{\hole_1, \hole_2} $ necessary to be consumed to simulate a source term step is $ \mu $. Hence, to allow the simulation of one step of $ R $, to disable the simulation of the respective other step of $ R $, and since $ \sigma'\left( \Encoding{\overline{a} + b + b.\success} \right) \equivAlpha \Encoding{\overline{b} + a + a.\success} $, there is some capability in $ \Encoding{\overline{a} + b + b.\success} $ as well as in $ \Encoding{\overline{b} + a + a.\success} $ and, to simulate a source term step, both of these capabilities have to be consumed. Moreover, since $ \sigma'\left( \Encoding{\overline{a} + b + b.\success} \right) \equivAlpha \Encoding{\overline{b} + a + a.\success} $, both capabilities are of the same kind, i.e. both are either input prefixes or both are output prefixes. Note that there is no possibility in \piAsyn to consume two capabilities of the same kind within the same target term step. Then it can not be avoided that for each of the simulations of the two steps exactly one of these two capabilities is consumed. In this case, none of the simulations can be completed, i.e., there is some local deadlock.
	
	Considering $ \Encoding{Q} $, such a deadlock leads to a term $ T_Q $ with $ \Encoding{Q} \Longmapsto T_Q $ and, since none of the source term steps is simulated, no unguarded occurrence of $ \success $ is reached, i.e. $ T_Q \not\Downarrow $. With that, such a deadlock leads to a contradiction.
	
	The only way to circumvent a deadlock in this situation, is that one of these capabilities is released by one of these simulations. To complete the simulation of this step later on, it has to be possible that the released capability is consumed again. But then it can not be avoided that this is done before the other simulation is finished, i.e. that leads back to the situation before. Then we introduce divergence, i.e., contradicts divergence reflection. Thus, it is not possible that the simulation of alternative source term steps is enabled concurrently.
\end{proof}
Note that, even if we allow separate choice within the target language, the simulation of the source term step on $ a $ can not immediately withdraw the encodings of the capabilities on $ b $ and vice versa. Because either we have to split each of these sums into an input and an output guarded sum or we have to convert each of them into a single sum with only separate choice. In the second case, since by compositionality both parameters have to be encoded in exactly the same way, we either result in two input guarded or two output guarded sums. But then we can not simulate a communication between these to sums within a single step and, moreover, we can not decide within a single step whether a considered capability can be used to successfully simulate a source term step. Unfortunately, the first step used to try whether we can use this capability to simulate a source term step removes all the other encoded capabilities of that sum, which violates operational correspondence. So Lemma \ref{lem:singleSimulation} holds as well for separate choice in the target language.

Now we can prove the main result of this paper.
\begin{thm} \label{thm:encodingSynchronyIntroduceCausality}
	Any good encoding from \piMix into \piAsyn introduces additional causal dependencies.
\end{thm}
By Lemma \ref{lem:contextInteract} and Lemma \ref{lem:singleSimulation} the context must contain some capability, i.e. some kind of lock, that must be consumed to enable the simulation of a source term step. Moreover, to enable the simulation of subsequent source term steps, the capability that was consumed from the context must be restored. Then, a subsequent simulating sequence is enabled by the consumption of a capability that was produced by a former simulating sequence; this interplay imposes a causal dependence between subsequent simulations of source term steps.
\begin{proof}
	By Lemma \ref{lem:contextInteract} and Lemma \ref{lem:singleSimulation} the context $ \ContextABPar{\hole_1, \hole_2} $ provides exactly one instance of $ \mu $, i.e., one capability, that needs to be consumed to simulate a source term step. Since $ \Encoding{Q} \Longmapsto Q_1' $, $ \Encoding{Q} \Longmapsto Q_2' $, and $ \Encoding{Q_1} \Longmapsto Q_2'' $, where $ Q_i' \asymp \Encoding{Q_i} $ for $ i \in \Set{ 1, 2 } $ and $ Q_2'' \asymp \Encoding{Q_2} $, i.e. $ \Encoding{Q} $ simulates two subsequently source term steps, the instance of $ \mu $ consumed by the simulation of the first source term step has to be restored during this first simulation such that the second step can be simulated. Thus, the simulation of the second step has to consume some capability $ \mu $ produced by the simulation of the first step. Then the simulation of the second step causally depends on the simulation of the first step, although any pair of subsequent steps of $ Q $ are causal independent. We conclude that the encoding function adds additional causal dependencies.
\end{proof}

\subsection{Discussion} \label{sec:discussion}

It is no fortuity that the counterexamples in the proof of Theorem \ref{thm:encodingSynchronyIntroduceCausality} rely on mixed choices. It is the power of mixed choice that allows \piMix to break initial symmetries. Note that the separation results of \cite{palamidessi03} and \cite{petersNestmann10} are based on that absolute difference in the expressive power of \piMix compared to $ \pi $-calculus variants without mixed choice. In \cite{gorla10}, the role of breaking symmetries for the separation result is not equally obvious. Nevertheless, the counterexample used there also relies on mixed choices and their ability to break symmetries.
In summary, the difference in the expressive power of \piMix compared to \piAsyn essentially|if not exclusively|relies on the expressive power of mixed choice.

\paragraph{Synchrony versus Guarded Choice.}
\label{sec:synchr-vers-choice}

It is debatable in how far a discussion on synchrony versus asynchrony can be separated from a discussion of choice.  In fact, even from a pragmatic point of view within our model of distributed reactive systems, it cannot.  It is part of the nature of reactive systems|in our case: systems communicating via message-passing along channels|that agents do not only listen to one channel at a time; they concurrently listen to a whole selection of channels. In this respect, as soon as a calculus offers a synchronous (blocking) input primitive, it is natural to extend this primitive to an input-guarded choice. Having mutual exclusion on concurrently enabled inputs is useful when thinking of a process's local state that may be influenced differently by any received information along the competing input channels.  (\emph{Joint input}~\cite{nestmann:joint-input}, as motivated in the join calculus \cite{fournet.gonthier:reflexive}, represents another natural and interesting generalisation.) Likewise, as soon as a calculus offers synchronous output, one may generalise this primitive to output-guarded choice.  This generalisation seems less natural, though, as the process's state would hardly be influenced by a continuation of one of the branches after an output.  However, having both input- and output-guards in the calculus, mixed choice becomes expressible.  Mixed choice is again also natural, as the successful execution of an output may prevent a competing input, including the effect of the latter on the local state.  These pragmatic arguments support the point of view that, in a message-passing scenario, any discussion of synchronous versus asynchronous interaction must consider a competitive context, as expressed by means of choice operators.

\paragraph{The Role of Mixed Choice.}
\label{sec:role-mixed-choice}

In the proof of Lemma \ref{lem:contextInteract} a counterexample based on mixed choice is used not only to rule out that the parallel operator is translated homomorphically (compare to the argumentation of the separation results in \cite{palamidessi03}, \cite{gorla10}, and \cite{petersNestmann10}) but also to prove that, in order to break symmetries, the context introduced to encode the parallel operator must interact with the encodings of its parameters. Moreover, only with mixed choice it is possible to give an example of a symmetric network with two conflicting steps on two different channels. This is necessary in the proof of Lemma \ref{lem:singleSimulation} to show that the context can not enable the simulation of more than one source term step at a time. Since the two source term steps are on different channels, their simulation is on the encoding of different capabilities, i.e. the simulations do not interfere by interaction on the same encoded capabilities. Because the two source term steps are in conflict, operational completeness in combination with success sensitiveness forbids that both simulations are completed successfully. We conclude that the simulation of one source term step must inevitably disable the successful completion of the conflicting source term step by the consumption of some necessary capability. Since the source terms are on different channels, this capability is not due to the encoding of a source term capability but added to the encoding of the corresponding process of the source term network. The symmetry of the source term allows to apply the name invariance criterion to conclude that both processes of that symmetric source term net are encoded symmetrically. Hence, there is one instance of the respective capability in each encoded process. Now, the simulation of one of the two conflicting steps has to consume both such capabilities. As explained above two concurrently enabled simulations then compete for these capabilities, which leads to deadlock or divergence. Note that deadlock would violate operational soundness.

Interestingly, the necessity of such a capability to rule out conflicting steps as well as the associated danger to introduce deadlock or divergence was already pointed out by Nestmann \cite{nestmann00}, presenting a good encoding from \piSep into~\piAsyn. To rule out conflicting steps on the same sum, the encoding of a sum introduces a so-called sum lock: a boolean-valued lock that is initially instantiated with true, but is set to false as soon as some summand of the sum commits to successfully simulate a source term step. Using this sum lock, the encoding function does not only forbid to use that summand twice but also to use any other summand of that sum to simulate subsequently source term steps. Hence, this sum lock of the encoding in \cite{nestmann00} does exactly occupy the role of the capability consumed by the simulation of a step to rule out the simulation of a conflicting step as described above. Note that the encoding presented in \cite{nestmann00} translates the parallel operator homomorphically and thus is no good encoding from \piMix into \piAsyn. \cite{nestmann00} explains that the application of that encoding function to terms with mixed choices|or, more precisely, in the presence of mixed choices with cyclic dependencies between the links of matching capabilities as in $ R $ or $ Q $|leads to exactly the deadlocked situation described in the proof above.

The proof of Theorem \ref{thm:encodingSynchronyIntroduceCausality} reveals the solution to circumvent this problem with the so-called cyclic sums. The encoding of the parallel operator has to ensure that there is at most one simulation of a source term step between the two parameters of that parallel operator at the same time. In fact, it is exactly this required blocking of alternative steps that leads to additional causal dependencies. As claimed in \cite{petersNestmann11}, the introduction of a lock at the level of the encoding of a parallel operator indeed suffices to circumvent the problem of cyclic sums. Note that, in the appendix of \cite{petersNestmann11}, the main line of argumentation of a respective proof is given. Moreover, it is explained in more detail how|in the context of that particular encoding|this lock leads to additional causal dependencies. It is the temporal blocking of the simulation of source term steps, necessary to avoid deadlock or divergence in case of conflicting source term steps, that leads to additional causal dependencies in case of concurrent source term steps.

\paragraph{True Concurrency.}

Let us take a closer look at the kind of additional causal dependencies an encoding function must, according to the above proof, introduce to encode synchronous interactions. We observe that the proof induces a causal dependency between the simulation of all subsequent source term steps between the two sides of the same parallel operator. Moreover, we observe that a later such step is causally dependent of a former one, but that no fixed order on the simulations of steps is induced. The same observation holds for the encoding presented in \cite{petersNestmann11}: it is forced to block some simulations of source term steps until another simulation is finished, but there is no directive on which kind of steps have to be simulated first. Hence, the simulation of two concurrent source term steps can still appear in either order but, if both steps are due to a communication over the same parallel operator, i.e. between the same processes of a network, the corresponding simulations can not be performed truly concurrently.
Note that the proof above does not state that the described additional causal dependency is the only kind of causal dependency any good encoding of synchrony has to introduce. This might be an interesting question for further research.

\section{Conclusion} \label{sec:conclusion}

We show that, in the context of the $\pi$-calculus, any good encoding of synchronous interaction within a purely asynchronous setting introduces additional causal dependencies and, thus, reduces the number of truly parallel steps of the original term. Moreover, the proof of this result further illustrates the importance of the role of mixed choice to distinguish the expressive power of \piMix and \piAsyn. To tighten that view on the importance of mixed choice for the introduction of causal dependencies in encoding synchrony, we have to show that the encoding from \piSep into \piAsyn presented in \cite{nestmann00} does not introduce additional causal dependencies. This statement seems intuitively correct; we leave its formal proof for further research.

As already mentioned, there is a companion paper \cite{schickePetersGoltz11} that proves a result similar to Theorem \ref{thm:encodingSynchronyIntroduceCausality} in the context of Petri nets. More precisely, they show that it is not always possible to find a finite, 1-safe, distributed net which is completed pomset trace equivalent to a given net. Note that completed pomset trace equivalence is sensitive to (local) deadlocks and causal dependencies. Hence, the connection between synchronous interactions and causalities, i.e., that any good encoding of synchrony changes the causal semantics of the source, is no effect of the representation of concurrent systems in either the $ \pi $-calculus or Petri nets, but seems to be a phenomenon of synchronous and asynchronous interactions in general.

\begin{figure}
  \begin{center}
	\tikzstyle{place}=[circle,draw=black,thick,minimum size=5mm]
	\tikzstyle{transition}=[rectangle,draw=black,thick,minimum size=5mm]
	\begin{tikzpicture}
		\node[place,tokens=1]	(p) at (2, 1.2) {};
		\node[place,tokens=1]	(q) at (4, 1.2) {};
		\node[transition]		(a) at (1, 0) {$ a $};
		\node[transition]		(b) at (3, 0) {$ b $};
		\node[transition]		(c) at (5, 0) {$ c $};
	   	
		\draw[->] (p) -- (a);
		\draw[->] (p) -- (b);
		\draw[->] (q) -- (b);
		\draw[->] (q) -- (c);
	\end{tikzpicture}
	\end{center}
	\caption{A fully reached, pure \textbf{M} \cite{schickePetersGoltz11}.} \label{fig:pureM}
\end{figure}
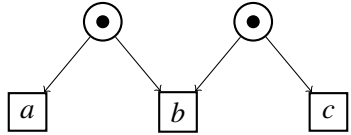

A closer look at the proof in \cite{schickePetersGoltz11} reveals that this proof depends on a counterexample including a so-called \emph{fully reached pure} \textbf{M} (see Figure~\ref{fig:pureM}). Similarly, our result depends on counterexamples that are symmetric networks including mixed choices. In both cases, the counterexample refers to a situation in the synchronous setting in which there are two distinct but conflicting steps. To solve this conflict, two simultaneous activities are necessary|in case of the $ \pi $-calculus the reduction of two sums, in case of Petri nets the removal of two tokens. In the asynchronous setting, this simultaneous solution must be serialised, e.g., by means of some kind of lock. It blocks the enabling of the asynchronous simulations of source term steps, such that no two simulations of conflicting source steps are enabled concurrently. In both formalisms, Petri nets and the $ \pi $-calculus, it is this temporal blocking of the simulation of source term steps|necessary to avoid deadlock or divergence in case of conflicting source term steps|that leads to additional causal dependencies.

However, apart from this apparent similarity, the relation between the two results leaves us with a number of open problems and incites further directions of research. To begin with, the requirements imposed on $\pi$-calculus implementations and Petri net implementations take rather different forms. Additionally, in contrast to the Petri net result, the present paper has no need to employ a (pomset based) equivalence to compare source and target terms and also does not need to deal with infinite implementations specifically. On the other hand, the Petri net result does not impose any restrictions on the encoding itself, but it connected source and target nets by means of behaviour only without any reference to the net structure. Finally, the Petri nets considered in \cite{schickePetersGoltz11} are not Turing-complete. So is it possible to derive the same result considering a Turing-complete formalism as for instance Petri nets with inhibitor arcs?
We hope to answer some of these questions in future work.

\addcontentsline{toc}{section}{References}
\bibliographystyle{eptcs}
\bibliography{SynchronyCausalityPi.bib}

\end{document}